\documentclass[final,3p,times]{elsarticle}

\usepackage{amssymb}
\usepackage{amsmath}

\usepackage{verbatim}
\usepackage{hyperref}

\begin{document}

\begin{frontmatter}

\title{Comment on "Beam blow up due to beamstrahlung in circular $e^+e^-$ colliders"}

\author{D. Shatilov} 
\ead{dshatilov@niu.edu}

\affiliation{organization={Northern Illinois University}, 
            addressline={1425 W. Lincoln Hwy.}, 
            city={DeKalb},
            postcode={60115}, 
            state={IL},
            country={USA}}

\begin{abstract}
The authors of the original article, \href{https://doi.org/10.1140/epjp/s13360-021-01485-x}{Valdivia Garcia and Zimmermann (2021)}, state that they "derive analytical formulae for the equilibrium beam parameters, considering the varying electromagnetic field during the collision". However, to calculate the bending radius in the field of the counter-propagating bunch, they employed an approximation that is completely unjustified. Consequently, a significant portion of the formulae derived in the article bears a very tenuous relationship to reality. For practical purposes, we propose a different solution -- one that is far simpler and more accurate.
\end{abstract}

\begin{keyword}
FCC-ee \sep CEPC \sep Beamstrahlung

\end{keyword}
\end{frontmatter}


\section{Hourglass effect and the definition of $\bar{y}$}

All notations used here and subsequently, as well as page and equation numbers, refer to the original article~\cite{Valdivia_Zimmermann}. In Eq.~(5), the authors introduced the parameter $G_y(\bar{s}, \bar{x}, \varPhi)$ that accounts for the hourglass effect, but also includes the effect of crab waist. It should be noted that the latter causes the beam's transverse profile to become significantly non-Gaussian, rendering Eq. (8) (the well-known Bassetti-Erskine formula) inapplicable. In general, no analytical expression for the beam-beam force $\mathcal{F}$ exists for such a beam profile; however, in simulations, this issue can be addressed using specialized grids to approximate $\mathcal{F}$~\cite{Shatilov_cw}. On the other hand, the impact of the crabbed beam profile on beamstrahlung is quite insignificant. Therefore, given the objectives of~\cite{Valdivia_Zimmermann}, it would be entirely logical to disregard it, thereby making Eq. (8) valid. In this case, $G_y$ depends only on $\bar{s}$.

Matters become even simpler thereafter, as the authors consider only the symmetric case, in which the dependence of $\sigma_y$ on the azimuth is the same for both colliding beams. Both the particle distribution function and the formula for $\mathcal{F}$ use the same ratio of $y$ to the vertical beam size at the given azimuth $\sigma_y(s)$. Therefore, it would be much more convenient to introduce the variable \hbox{$\bar{y} = y/\sigma_y(s)$} instead of $\bar{y} = y/\sigma_y^*$. Then, one simply needs to replace $\sigma_y^*$ with $\sigma_y$ in Eq. (8) and remove the factor $G_y$ from all equations, which significantly simplifies them. An explicit dependence on $\sigma_y$ remains only in Eqs.~(8) and (9), but for flat beams it vanishes. This reflects a long-known property: for flat beams, beamstrahlung does not depend on $\sigma_y$. Consequently, there will be no dependence on the hourglass effect either.

\section{Approximation for the function $\mathcal{F}$}

 On page 6, after Eq.~(12), it is written: "assuming $|\bar{x}| \lesssim 1$ and $|\bar{y} / G_y| \lesssim 1$, which we justify by the Gaussian weighting of $\bar{x}$ and $\bar{y}$ in Eq.~(10), we expand the integrand as $exp(-\tau^2/2) \approx (1 - \tau^2/2 + ...) \approx 1$". This is a key phrase for all subsequent equations, and it raises a number of serious questions:

\begin{enumerate}
    \item The exponent in Eq.~(10) contains not $-\bar{x}^2/2$, but $-(\bar{x} + \bar{z}\varPhi)^2/2$. How did the authors conclude that $|\bar{x}| \lesssim 1$ can be justified by the Gaussian weighting of $\bar{x}$? For head-on collision, one might -- with some stretch -- agree with this. But what about the case where the Piwinski angle $\varPhi \gg 1$?
    \item Now let us take into account that $\bar{x}$ in the inequality $|\bar{x}| \lesssim 1$ should be regarded as the first argument of the function $\mathcal{F}$, and this is not at all the $\bar{x}$ that appears in Eq.~(10). On page 8, the notation $u \equiv (\bar{x} - 2 \varPhi \bar{s} + \varPhi \bar{z})$ is introduced for this argument; and further on in the text, it is precisely $u$ that is substituted into the approximation formula (14). But what are the grounds for assuming $|u| \lesssim 1$? Here, it is necessary to take into account the dependence of $u$ on $\bar{s}$. The "Gaussian weight" equivalent corresponding to the azimuth is represented on the right-hand side of Eq.~(7) by a factor with the exponent $-(2\bar{s} - \bar{z})^2/2$. Accordingly, following the authors' logic, one can introduce the constraint $|2\bar{s} - \bar{z}| \lesssim 1$. If, for simplicity, we set $\bar{x}=0$, this leads to $|u| \lesssim \varPhi$. However, it would be incorrect to interpret this inequality literally, especially for small $\varPhi$. Its essence is that $|u|$ {\it can reach} values of $\varPhi$, and this depends primarily on the azimuth rather than the coordinates, meaning it applies to all particles. Now, let us recall that we are specifically interested in the case $\varPhi \gg 1$.
    \item The next question is about the expansion $exp(-\tau^2/2) \approx (1 - \tau^2/2 + ...) \approx 1$". Even if one accepts the aforementioned assumption $|\tau| \lesssim 1$, it remains completely unclear why the dependence on $\tau$ vanished as a result. It is evident that the authors' aim to impose a restriction on $|u|$ and $|\bar{y}|$ is driven precisely by the desire to reduce the expansion of this exponential to just the first (constant) term. However, this would require a stricter constraint~--~at least $|\tau| \lesssim 0.7$.
    \item After Eq.~(13), the authors write: "We will later compare all results obtained using this and the following approximation with those obtained by numerically computing the original complete expression (8), finding differences at the few percent level". And later, after Eq.~(14): "In the following, we compute various integrals involving the function $\mathcal{F}$ either using its fully general form, Eq.~(8), or using the approximation $\mathcal{F}_{approx}$ of (14), that is much faster to evaluate, and compare the respective results". However, they made some comparisons only for a head-on collision, and only for $\int ds \, \langle 1 / \rho \rangle$. Why, then, are there no comparisons for $\int ds \, \langle 1 / \rho^3 \rangle$ and for the case $\varPhi \gg 1$? Did the authors consider it unnecessary? Or, for some reason, undesirable? We will try to fill this gap below.
\end{enumerate}

\begin{figure*}
\centering
\resizebox{1.0\textwidth}{!}{
\includegraphics{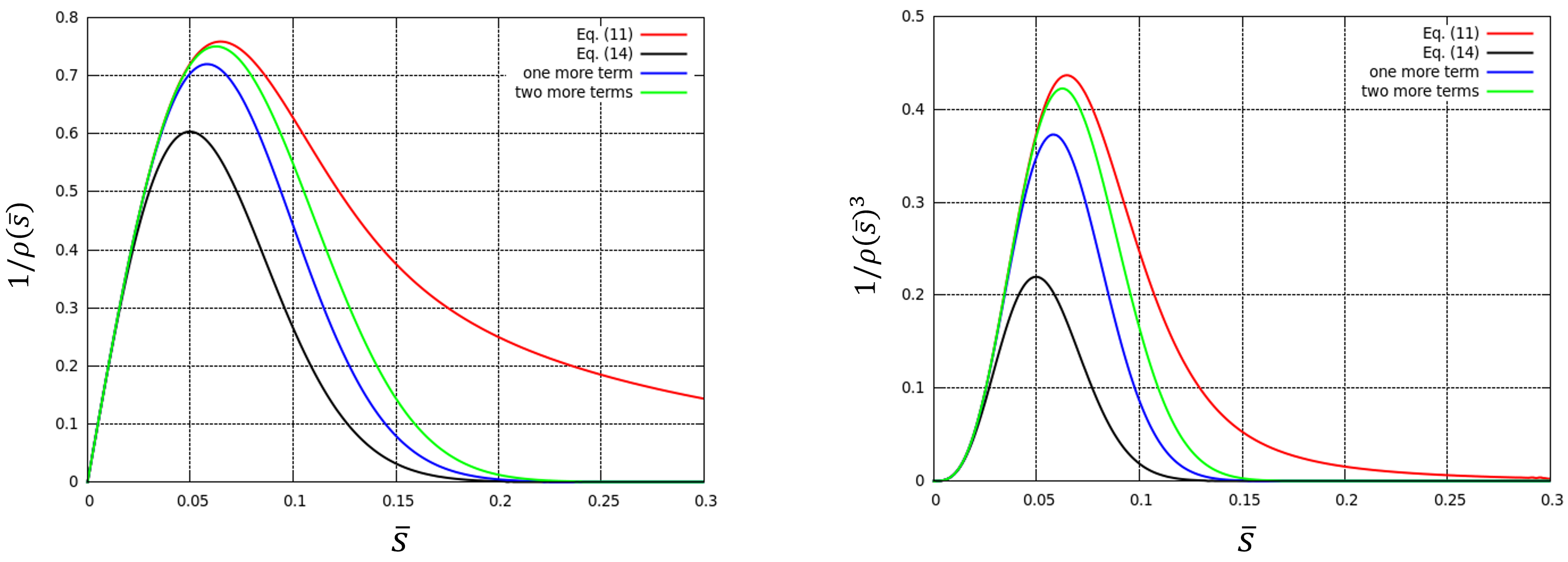}}
\caption {$1 / \rho(\bar{s})$ (left) and $1 / \rho(\bar{s})^3$ (right) for $\varPhi = 10$, $\bar{x} =  \bar{y} = \bar{z} = 0$. The horizontal axis is the "normalized azimuth", $\bar{s} = 0.05$ corresponds to $|u| = 1$. The values on the vertical axis are expressed in "arbitrary units", as we did not take constant factors into account. For negative $\bar{s}$, the pictures will be symmetrical.}
\label{Fig:bunch_center}
\end{figure*}

Let us start with a simple case: flat beams, $\varPhi = 10$, and a particle at the center of bunch, where the density is maximal. In this case, $|u| = 20|\bar{s}|$. The graphs of the functions $1 / \rho(\bar{s})$ and $1 / \rho(\bar{s})^3$ for this particle are shown in Fig.~\ref{Fig:bunch_center}. The red curve corresponds to the function $\mathcal{F}$ calculated using Eq.~(11). The black curve corresponds to approximation~(14). The blue and green curves correspond to approximations in which one and two additional terms have been retained in the expansion of $exp(-\tau^2/2)$. This example illustrates the approximation error for the horizontal component of $\mathcal{F}$. The area size under the curves represent the contribution to the corresponding radiation integrals, whereas approximation (14) underestimates them several-fold.  

\begin{figure*}
\centering
\resizebox{1.0\textwidth}{!}{
\includegraphics{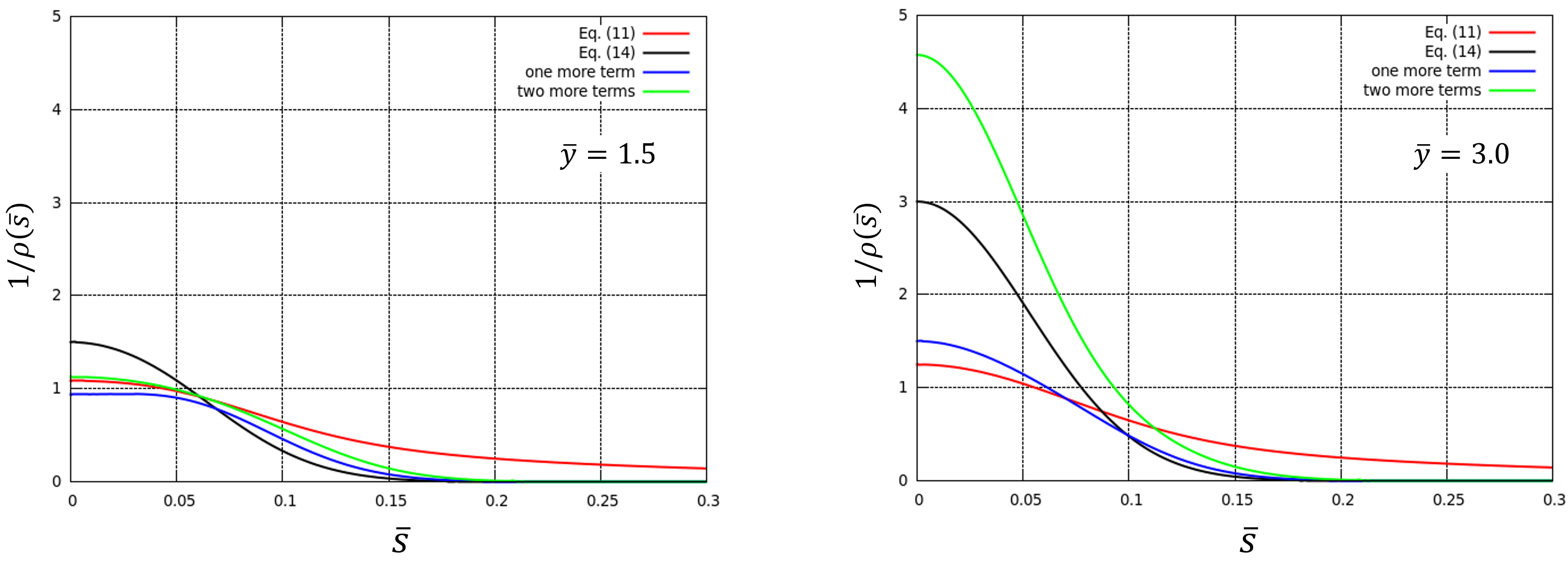}}
\caption {$1 / \rho(\bar{s})$ for $\varPhi = 10$, $\bar{x} =  \bar{z} = 0$ and two values of $\bar{y}$.}
\label{Fig:vert_coord}
\end{figure*}

\begin{figure*}
\centering
\resizebox{1.0\textwidth}{!}{
\includegraphics{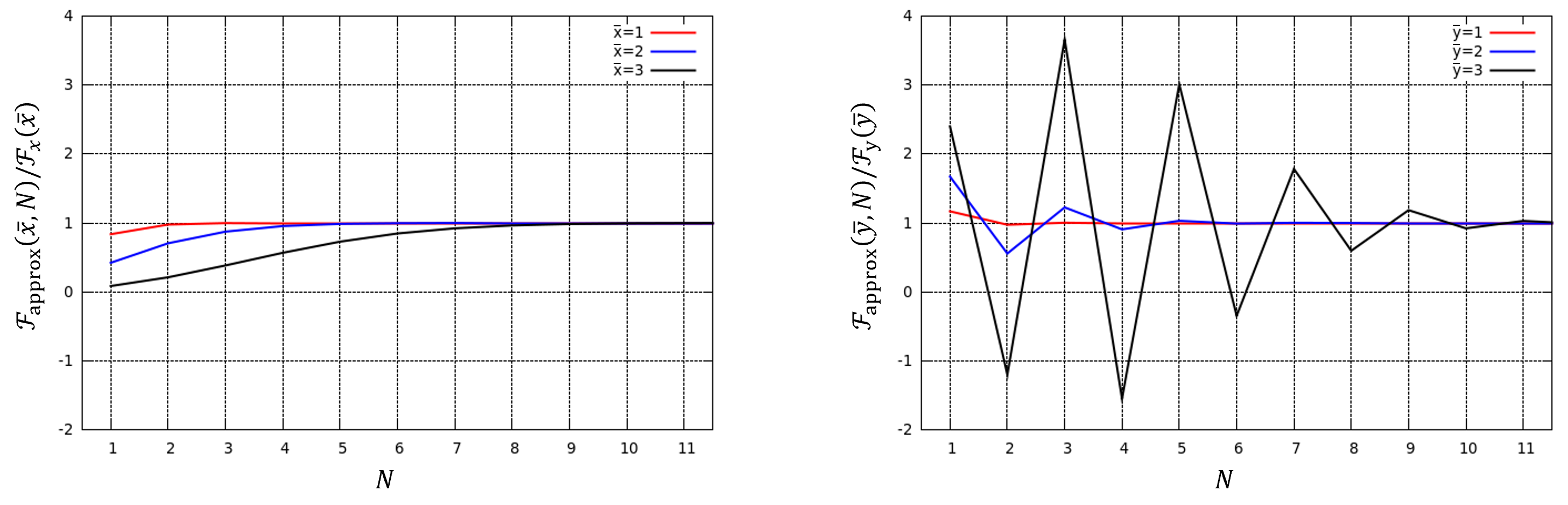}}
\caption {Convergence of the horizontal (left) and vertical (right) parts of $\mathcal{F}_{approx}$ as a function of the corresponding normalized coordinate and the number of terms $N$ in the expansion of $exp(-\tau^2/2)$ in Eq.~(11).}
\label{Fig:converge}
\end{figure*}

Let us now consider the approximation error associated with the vertical component of $\mathcal{F}$. Fig.~\ref{Fig:vert_coord} shows plots of $1 / \rho(\bar{s})$ for particles with the vertical coordinates $\bar{y} = 1.5$ and $3$. For flat beams, the dependence of $\mathcal{F}_y$ on $\bar{y}$ becomes weak when $\bar{y} > 1$, which is also evident when comparing the red curves in the left and right panels. However, approximation~(14) (black curve) exhibits different behavior, and for $1 / \rho(\bar{s})^3$ it will be even more noticeable. It should also be noted that for $\bar{y} = 1.5$, including each subsequent term of the exponential expansion reduces the discrepancy, whereas for $\bar{y} = 3$, the addition of the third term (green curve) leads to a significant increase in it. For larger values of $\bar{y}$, the situation worsens further: even the second term can increase the error. This is due to the fact that the convergence behavior of the series for $\mathcal{F}_x$ and $\mathcal{F}_y$ differs significantly at large values of the argument, as shown in Fig.~\ref{Fig:converge}.

It is worth noting that the improvement in approximation quality observed when the second term of the exponential expansion is included for the case $\bar{y} = 3$ (as seen in the right panel of Fig.~\ref{Fig:vert_coord}, where the blue line runs closer to the red line than the black one) is misleading. As shown in the right panel of Fig.~\ref{Fig:converge}, the quantity $\mathcal{F}_{approx}(\bar{y})$ changes sign in this case ($N = 2$), yet its absolute value is close to the true value. Since the absolute value is what matters for the bending radius, we observe reasonably good agreement in Fig.~\ref{Fig:vert_coord}; nevertheless, an approximation that yields the wrong sign can hardly be considered good.

On the other hand, the contribution of large $\bar{y}$-values to the integrals is indeed suppressed by the Gaussian exponential; however, if the discrepancy is large, this effect may be noticeable. To gain a more general understanding of the problem, it is necessary to shift from considering individual particles to the entire beam.

\section{Radiation integrals}

As shown above, the approximation errors for $\mathcal{F}_x$ and $\mathcal{F}_y$ exhibit a number of significant differences. In the case of $\mathcal{F}_x$, large values of the argument ($|\bar{u}| > 1$) are primarily associated with a large Piwinski angle ($\varPhi \gg 1$) and affect all beam particles, regardless of their coordinates relative to the center of bunch. For $|\bar{u}| > 1$, the approximation significantly underestimates the value of $\mathcal{F}_x$. Including one or two additional terms of the exponential expansion yields some improvement, but this is clearly insufficient. In the case of $\mathcal{F}_y$, large values of the argument ($|\bar{y}| > 1$) are not linked to $\varPhi$, and the fraction of such particles is limited by the Gaussian distribution. For $|\bar{y}| > 1$, the approximation significantly overestimates the value of $\mathcal{F}_y$; moreover, including a few (not many) additional terms of the exponential expansion can actually exacerbate the problem. 

\begin{figure*}
\centering
\resizebox{1.0\textwidth}{!}{
\includegraphics{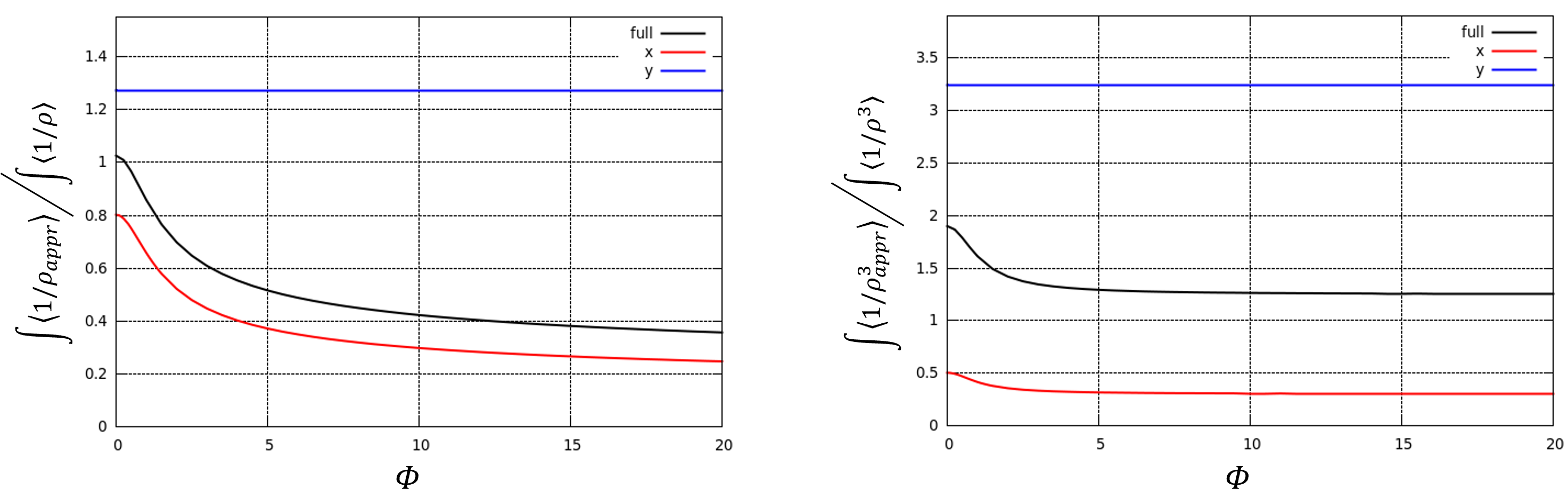}}
\caption {The ratios between values obtained using approximation (14) and the full analytical solution based on Eq. (11) for $\int ds \, \langle 1/\rho \rangle$ (left) and $\int ds \, \langle 1/\rho^3 \rangle$ (right), as a function of $\varPhi$. The red and blue curves correspond to the bending radius in the horizontal ($\rho => \rho_x$) and vertical ($\rho => \rho_y$) planes only.}
\label{Fig:integrals}
\end{figure*}

As follows from Eq.~(7), $1/\rho \propto |\mathcal{F}| = \sqrt{\mathcal{F}_x^2 + \mathcal{F}_y^2}$; and we already know that horizontal and vertical approximation errors partially compensate for each other. Therefore, to better understand the quality of the approximation, alongside $1/\rho$, we also consider $1/\rho_x \propto |\mathcal{F}_x|$ and $1/\rho_y \propto |\mathcal{F}_y|$. Fig.~\ref{Fig:integrals} shows the ratios between values obtained using approximation (14) and the full analytical solution based on Eq. (11) for $\int ds \, \langle 1/\rho \rangle$ and $\int ds \, \langle 1/\rho^3 \rangle$ (full and x,y-parts), as a function of $\varPhi$. As expected, the y-ratio (blue line) does not depend on $\varPhi$, whereas the x-ratio (red curve) -- and, consequently, the full ratio (black curve) -- decreases as $\varPhi$ increases.

As shown by the black curve on the left side of Fig.~\ref{Fig:integrals}), the compensation of the two errors is nearly perfect for $\int ds \, \langle 1/\rho \rangle$ in the case of head-on collision ($\varPhi$ = 0). This is the only point exhibiting such a small error -- and it is precisely this case that is examined in detail in Section 5 of~\cite{Valdivia_Zimmermann}. This integral does not appear later in~\cite{Valdivia_Zimmermann}, as it is not required to obtain the results presented in the article.

We are primarily interested in $\int ds \, \langle 1/\rho^3 \rangle$ for $\varPhi \gg 1$. As we can see, approximation (14) reduces the contribution of the horizontal component by more than a factor of three and increases the contribution of the vertical component by more than a factor of three. This is precisely what characterizes the accuracy of this approximation. However, due to the mutual compensation of these two errors, the value of the full integral increases by a factor of only \hbox{$f_{appr} \approx 1.26$}. And how will this affect the equilibrium energy spread? For simplicity, let us consider the FCC-ee Z-pole case, where $\sigma_{\delta}$ increases by approximately 3.5 times. In this case, the $\sigma_{\delta, \, SR}^2$ term in Eq.~(30) can be neglected, yielding $\sigma_{\delta}^4 \propto V$. However, one must also take into account that $V \propto 1/\varPhi$ (see Eqs.~(25) and (35)) and $\varPhi \propto \sigma_z \propto \sigma_{\delta}$. Ultimately, we find that $\sigma_{\delta} \propto f_{appr}^{1/5}$, meaning the error is less than $5 \, \%$.

If the $\sigma_{\delta, \, SR}^2$ term in Eq.~(30) is taken into account, the error becomes even smaller. This is especially true for higher energies, where the enhancement factor for $\sigma_{\delta}$ is significantly smaller because the relative contribution of $\sigma_{\delta, \, SR}$ to $\sigma_{\delta}$ is larger. This example demonstrates how little the final result depends on the quality of the approximation -- though this does not excuse its inadequacy. And one more detail: approximation~(14) overestimates the value of the radiation integral $\int ds \, \langle 1/\rho^3 \rangle$, whereas \hbox{Figs. 3-5} in~\cite{Valdivia_Zimmermann} show the opposite effect. How did this come about? And what exactly did the authors depict in these figures?

As we have seen, approximation~(14) is poor, and improving it is highly problematic -- not to mention the difficulty of evaluating subsequent integrals to express the result in a usable form. On the other hand, perhaps it is better to have at least some approximation than none at all? At this point, we must address an important question: what is the actual purpose of this? If the goal is to meet practical needs, the problem becomes much simpler to solve. As the authors of~\cite{Valdivia_Zimmermann} rightly noted, for flat beams (and no other colliders where beamstrahlung would play a significant role are currently envisaged), the integrals of interest depend on a single parameter: $\varPhi$. Thus, one simply needs to evaluate them numerically and construct a table for various values of $\varPhi$ -- see Table~\ref{Table_1}, which presents three dimensionless integrals that must be divided by 100 and multiplied by the dimensional coefficients $Q_n$ to obtain the actual values. The factor of 100 was chosen solely for the sake of tabular presentation, so that three significant digits fit into a narrow field. The coefficients $Q_n$ are the same as those in~\cite{Valdivia_Zimmermann} -- see Eqs.~(16) and (21); however, the factor $(2\pi)^{-3/2}$ in Eq. (21) should be removed. Simple interpolation using Table~\ref{Table_1} ensures an accuracy better than one percent, eliminating the need for complex formulae involving elliptic integrals.

\begin{table}[t]
\centering
{\footnotesize
\begin{tabular}{c | c c c c c c c c c c c c c c}
  $\varPhi$                  &  0   & 0.25 & 0.5  & 0.75 &  1   & 1.5  &  2   & 2.5  &  3   &  3.5 &  4   &  4.5 &  5   &  6  \\ \hline
  $\int ds \, \langle 1/\rho \rangle$   & 93.1 & 92.8 & 91.6 & 89.1 & 85.5 & 77.3 & 69.4 & 62.6 & 57.0 & 52.3 & 48.3 & 44.9 & 42.0 & 37.3 \\
  $\int ds \, \langle 1/\rho^2 \rangle$ & 53.6 & 53.3 & 52.2 & 50.3 & 47.7 & 41.8 & 36.4 & 31.8 & 28.0 & 25.0 & 22.5 & 20.5 & 18.8 & 16.0 \\
  $\int ds \, \langle 1/\rho^3 \rangle$ & 37.8 & 37.5 & 36.5 & 34.8 & 32.7 & 28.2 & 24.1 & 20.7 & 18.1 & 16.0 & 14.3 & 12.9 & 11.7 & 9.92 \\
\end{tabular}}

\vspace{0.4cm}

{\footnotesize
\begin{tabular}{c | c c c c c c c c c c c c c c}
  $\varPhi$                  &  7   &  8   &  9   &  10  &  12  &  14  &  16  &  18  &  20  &  22  &  24  &  26  &  28  &  30  \\ \hline
  $\int ds \, \langle 1/\rho \rangle$   & 33.6 & 30.6 & 28.2 & 26.1 & 22.9 & 20.4 & 18.4 & 16.8 & 15.5 & 14.4 & 13.5 & 12.7 & 12.0 & 11.3 \\
  $\int ds \, \langle 1/\rho^2 \rangle$ & 14.0 & 12.4 & 11.1 & 10.1 & 8.49 & 7.34 & 6.45 & 5.76 & 5.20 & 4.74 & 4.36 & 4.03 & 3.75 & 3.50 \\
  $\int ds \, \langle 1/\rho^3 \rangle$ & 8.59 & 7.56 & 6.75 & 6.10 & 5.11 & 4.39 & 3.85 & 3.42 & 3.09 & 2.81 & 2.57 & 2.38 & 2.21 & 2.06 \\
\end{tabular}}

\caption{Radiation integrals without factors $Q_n / 100$ for various values of $\varPhi$.}
\label{Table_1}
\end{table}

\section*{Acknowledgements}
This manuscript has been authored by Fermi Forward Discovery Group, LLC under Contract No. 89243024CSC000002 with the U.S. Department of Energy, Office of Science, Office of High Energy Physics.

\end{document}